# Uma Contribuição à Construção de Espelhos Parabólicos

*(A Contribution for the Construction of Parabolic Mirrors)*

Leandro Aparecido Nogueira de Paula,[1] Pedro Raggio[2] e André Koch Torres Assis[3]

Instituto de Física `Gleb Wataghin´
Universidade Estadual de Campinas – Unicamp
13083-970 Campinas, São Paulo

**Resumo:** Apresentamos um novo procedimento para a construção de espelhos parabólicos utilizando materiais de baixo custo. Construímos um sistema girante composto por fios de náilon, anzóis de pesca e uma bacia de plástico. Colocamos gesso líquido na bacia e a colocamos em rotação constante em relação à terra. Um líquido adquire um perfil parabólico ao girar com uma velocidade angular constante em um referencial inercial na presença de um campo gravitacional uniforme vertical. Ao mantermos o gesso girando por um longo tempo, ele solidifica no formato parabólico. Usamos este paraboloide de gesso solidificado como modelo para construir um contra-molde de fibra de vidro e resina. Sobre este contra-molde esticamos papel laminado e em seguida despejamos gesso pastoso sobre ele. Com isto obtém-se um espelho parabólico feito de papel laminado e gesso. Nosso objetivo aqui é apenas o de apresentar um procedimento novo para a construção de espelhos parabólicos utilizando materiais de baixo custo. Isto permite que este procedimento seja explorado por professores e alunos de ensino médio e universitário.

**Palavras-chave:** Espelho parabólico, materiais de baixo custo.

**Abstract:** We present a new procedure for the construction of parabolic mirrors using low cost materials. We build a spinning system composed of nylon threads, fish hooks and a plastic bucket. We pour liquid plaster into the bucket and set it in constant rotational motion relative to the earth. A liquid substance assumes a parabolic profile when spinning at constant angular velocity relative to an inertial frame under the influence of an uniform vertical gravitational field. By keeping the bucket under rotation for a long time, the plaster solidifies into a parabolic format. We utilize this solidified plaster paraboloid as a model to construct a counter-mould of glass fibre and resin. Over this counter-mould it is placed stretched laminated foil and then it is poured thick plaster over it. In this way it is obtained a parabolic mirror made of laminated foil and plaster. Our only objective here is to present a new method for the construction of parabolic mirror using low cost materials. This allows further exploration of this procedure by teachers and students either in high-schools or in universities.

---

[1] E-mail: leandroifgw@yahoo.com.br.
[2] E-mail: praggio@ifi.unicamp.br.
[3] Homepage: http://www.ifi.unicamp.br/~assis e e-mail: assis@ifi.unicamp.br.

## 1. Introdução

Os raios de luz paralelos ao eixo óptico de um espelho esférico ao incidirem sobre sua superfície refletora não são refletidos para um mesmo ponto. Isto gera o fenômeno denominado de aberração esférica. Este efeito, muitas vezes indesejado, pode ser eliminado se utilizamos um espelho parabólico. Mas na prática, a construção deste tipo de espelho é tediosa e cara uma vez que exige muitas horas de trabalho sendo preciso, às vezes, recorrer até a empresas especializadas (BERNARDES, 2006).

Neste artigo apresentamos um novo procedimento para a construção de espelhos parabólicos utilizando materiais de baixo custo. Seu pioneirismo se refere ao fato de termos solidificado o formato parabólico dos líquidos em rotação e ao construirmos um espelho parabólico com procedimento e materiais não usuais. Como o procedimento é inédito, muitos melhoramentos ainda podem ser feitos nas etapas que descreveremos. E de cada etapa muitos resultados diversos podem ser extraídos.

Construímos um sistema girante composto por fios de náilon, anzóis de pesca e uma bacia de plástico. Colocamos gesso líquido na bacia, que foi colocada em rotação constante em relação à terra. Um líquido adquire um perfil parabólico ao girar com uma velocidade angular constante em um referencial inercial na presença de um campo gravitacional uniforme vertical. Ao mantermos o gesso girando por um longo tempo, ele solidifica no formato parabólico. Usamos este paraboloide de gesso solidificado como modelo para construir um contra-molde de fibra de vidro e resina. Sobre este contra-molde esticamos papel laminado e em seguida despejamos gesso pastoso sobre ele. Com isto obtivemos um espelho parabólico composto de papel laminado e gesso.

## 2. Sistema Girante

Um sistema girante foi construído utilizando fios de náilon, anzóis de pesca e uma bacia de plástico. Com este material construímos três partes distintas: um cabo de sustentação composto de fios de náilon, um recipiente para colocar líquido (a bacia), e as alças feitas de náilon que ligam o recipiente ao cabo de sustentação, conforme mostrado na Fig. 1. O sistema assim constituído mostrou ter a propriedade de ser muito estável em rotação, girando aproximadamente durante um quarto de hora com uma velocidade angular constante em relação à terra. Isto permitiu a formação de uma superfície líquida parabólica muito precisa, formando até imagens muito definidas e nítidas com natureza de imagens de espelho parabólico. Por esta simplicidade de construção, estabilidade e efeitos derivados, este sistema girante pode ser considerado como um instrumento científico de boa precisão.

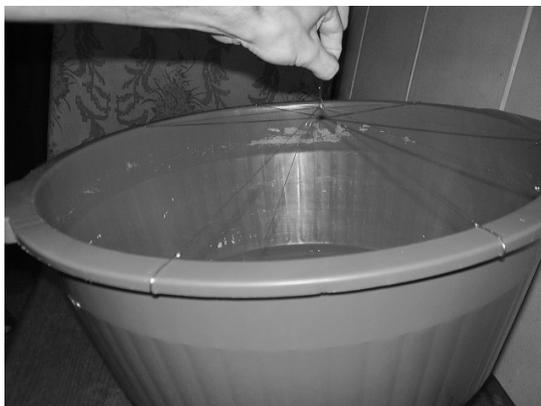

**Fig. 1:** Sistema girante suspenso pelas mãos ao invés do cabo de sustentação.

### 2.1. Cabo de sustentação

O cabo de sustentação da bacia é constituído por dez fios de náilon de aproximadamente 0,33 mm de espessura e 5 metros de comprimento cada um. Foi observado que com esta quantidade de fios e com este comprimento o cabo nunca se rompia qualquer que fosse o número de rotações (dentro de certos limites máximos). Ao mesmo tempo, o sistema acumulava uma boa quantidade de energia potencial de torção no cabo. O rompimento do cabo não ocorre neste caso porque os dez fios se entrelaçam impedindo que cada fio seja girado em torno de seu próprio eixo. Um único fio de náilon como o que foi utilizado aqui suporta muito bem aproximadamente doze quilogramas de tensões longitudinais, mas se rompe facilmente devido a tensões de cisalhamento ou de torção.

Há também uma diferença no comprimento entre o cabo torcido (com os fios entrelaçados) e o cabo relaxado (sem entrelaçamento). Pois quando torcemos o cabo, este vai se tornando cada vez mais curto em razão do entrelaçamento dos fios, atingindo um comprimento mínimo.

A descrição deste tipo de torção de entrelaçamento dificilmente se encontra nos livros didáticos e merece um estudo mais detalhado. Durante as experiências, aumentando o número de fios vem que as tensões longitudinais passaram a ter um papel mais importante para o acúmulo de energia potencial no cabo do que as tensões de cisalhamento. O cabo também passou a elevar o sistema à medida que a bacia era girada.

### 2.2. Alças

Há quatro alças de sustentação da bacia. Cada alça é constituída por dois fios de náilon com um anzol amarrado em cada extremidade. As quatro alças se cruzam na região central da bacia, onde foi colocado mais um anzol. Este anzol funcionou automaticamente como um ponto central para encaixar o cabo de sustentação e ao mesmo tempo como um dispositivo de ajuste fino para equilibrar o sistema. Por meio deste anzol as alças são suspensas pelo cabo de sustentação, formando então um formato octogonal, como que composto de oito aros.

Isto resultou em um sistema de alças, com os anzóis igualmente espaçados presos no recipiente, conforme mostra a Fig. 2. O cálculo para determinar o espaçamento dos anzóis ou o comprimento das linhas tracejadas fictícias é uma questão simples de geometria.

Quando o recipiente é preenchido com água, a região central de cruzamento das alças é puxada para cima pelo anzol que está no centro preso no cabo de sustentação. Com o recipiente que utilizamos foram necessários no mínimo sete quilogramas de água para equilibrar o sistema estaticamente.

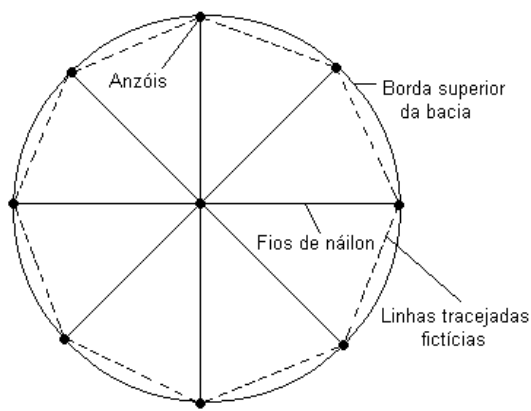

**Fig. 2:** Esquema octogonal das alças**.**

Uma quantidade menor de água colocada na bacia fazia a bacia pender para os lados, mesmo se o anzol central de ajuste fino estivesse rigidamente fixado no cruzamento das alças. Uma quantidade maior de água, mas não tão grande a ponto de romper algum fio, tornava o sistema mais estável.

Os principais responsáveis pela estabilidade do sistema girante foram a quantidade de líquido no recipiente e o sistema octogonal de alças. Entretanto, quando se usou gesso líquido ao invés de água, como será descrito na terceira seção, observou-se que não é somente a quantidade de líquido que é responsável pela estabilidade, mas também o volume de líquido.

### 2.3. Recipiente

Uma bacia de plástico de aproximadamente cinqüenta centímetros de diâmetro e vinte centímetros de altura, além de ser suficientemente rígida, precisa ter dobras laterais na sua extremidade superior para prender os anzóis. Esta bacia de plástico grande, rígida e com dobras na extremidade superior, foi utilizada como recipiente para fazer as experiências com água e gesso líquido e realizar a conseqüente solidificação do parabolóide de gesso.

### 3. Molde Parabólico

As imagens de corpos formadas pela reflexão na camada de água acima do gesso líquido em rotação foram usadas como indicadores de que a superfície era parabólica. Quando estas imagens estavam bem focalizadas isto significava que a superfície era parabólica.

O balde sem líquido no interior permanece em equilíbrio estático caso o cabo de sustentação não esteja torcido. Ao adicionar oito quilogramas de água no interior do recipiente este fica mais pesado e ainda continua em equilíbrio estático. Gira-se o recipiente com água colocando as pontas dos dedos em lados diametralmente opostos do recipiente e torcendo o cabo de sustentação. Com isto é fornecida energia potencial de torção ao sistema. O recipiente é girado com as mãos por aproximadamente dez minutos. Depois o sistema é parado e retira-se totalmente a água do seu interior. Quanto ao gesso líquido, antes de tudo, é necessário respeitar uma certa proporção entre a quantidade de água e o pó de gesso durante o seu preparo. A proporção em massa é de três partes de água para quatro partes de pó de gesso. Colocam-se então dez quilogramas e meio de gesso líquido no lugar da água. Esta etapa exige um pouco de cuidado, pois quando se está retirando a água, o recipiente vai sendo puxado para cima pelo cabo de sustentação torcido, e deve-se tomar cuidado para o anzol central não sair de sua posição que equilibra o sistema. E quando o gesso líquido é colocado no balde, o cabo vai sendo puxado para baixo pelo peso do recipiente e o mesmo cuidado deve ser tomado com relação ao anzol central. Depois que todo o gesso líquido for colocado na bacia, solta-se o recipiente para que ele entre em rotação automaticamente devido à energia potencial de torção do cabo.

Quando o sistema está estabilizado e em rotação constante as imagens reais formadas acima da superfície parabólica do gesso líquido são bem nítidas e definidas e, portanto, muito bem focalizadas. Assim podemos dizer que há um foco definido devido à forma parabólica do gesso líquido e à camada superior de água. Para manter este foco basta manter com a mesma qualidade e nitidez a imagem formada acima da superfície parabólica. A equação de uma secção parabólica em um plano de coordenadas $x$ e $z$ é dada por $z = (1/4p)x^2$, onde $p$ é a posição focal a partir do vértice da secção parabólica (SANTOS, 2004). E por outro lado temos que a mesma equação relacionando $x$ e $z$ é dada por $z = (g/2\omega^2)x^2$, onde $g$ é a aceleração da gravidade local e $\omega$ a velocidade de rotação da bacia em relação à Terra (ASSIS, 1998). Assim igualando estas duas equações temos que $p = g/2\omega^2$, que relaciona o foco com a velocidade de rotação do parabolóide. Quando a bacia começa a girar a velocidade de rotação começa a aumentar, embora com aceleração muito pequena. Isto modifica a estrutura da imagem, começando pela região central, aumentando em direção às bordas. Isso é resultante da distorção da forma parabólica do gesso líquido devida ao aumento na velocidade de rotação do sistema. Então quando a imagem começa a distorcer na região central, leva-se os dedos levemente nas laterais opostas do recipiente para frear o sistema e, assim, restabelecer a imagem anterior. Com isto também se restabelece o foco anterior, como é mostrado na última equação acima.

Alguns fatores são muito importantes para o sucesso da experiência: (I) o ponto de consistência do gesso líquido imediatamente antes do sistema entrar em rotação, (II) a velocidade constante de rotação do sistema em relação à terra, e (III) o tempo de duração que o sistema permanece girando, que é de aproximadamente uns vinte minutos no primeiro semi-período global (ou seja, enquanto o sistema está girando apenas em uma direção). Se o gesso líquido secar quando o sistema estiver girando tão rapidamente de modo que apareça o fundo do balde, então o paralóide sólido de gesso ficará sem fundo. Por outro lado, se o gesso líquido for pouco consistente antes de o sistema começar a girar, então não haverá tempo suficiente para solidificar o gesso. Se isto ocorrer o parabolóide se desmanchará quando o sistema parar. O ponto de consistência deve ser tal que o gesso se solidifique durante o primeiro semi-período global.

Após ter secado completamente, o molde parabólico de gesso é retirado do recipiente. Este processo de retirada é feito simplesmente virando o recipiente de cabeça para baixo. Como o gesso sólido não gruda na bacia, ele se solta com facilidade. O gesso sólido deve ser derrubado sobre uma almofada ou qualquer outra coisa macia para não danificá-lo. Em seguida, é necessário desbastar a parte superior do sólido para não ficar soltando pedacinhos de gesso. A Fig. 3 mostra o molde parabólico de gesso solidificado usando o procedimento de focalizar as imagens.

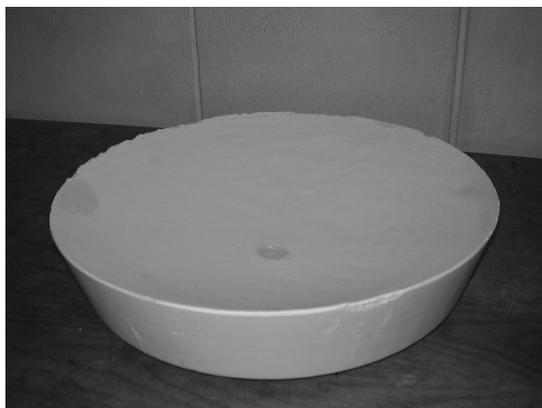

**Fig. 3**: Molde parabólico de gesso solidificado pelo método das imagens focalizadas.

## 4. Contra-Molde Parabólico

Um contra-molde de fibra de vidro e resina é construído utilizando o molde parabólico de gesso solidificado.

A fibra de vidro deve ser recortada e colocada sobre o molde parabólico de maneira que cubra toda a área parabólica. Deve-se tentar moldar com as mãos a fibra de vidro na concavidade parabólica amassando a fibra de vidro na direção desta concavidade. Espalha-se resina sobre a fibra com um pincel, começando pelo centro da concavidade e seguindo em direção à periferia. O processo de espalhar a resina deve ser rápido para que não haja a formação de bolhas. Deve-se pressionar o pincel molhado na resina contra as bolhas, caso sejam formadas. As regiões da fibra de vidro que se enrugam por excesso deste material devem ser removidas retirando este excesso ao puxar a fibra com as mãos até que se desprenda. A resina deve continuar sendo espalhada contra a própria fibra de vidro até que a fibra e a resina se assentem na concavidade tomando a forma parabólica. E espera-se secar.

Depois de secar por aproximadamente meia hora, o contra-molde de fibra de vidro e resina deve ser cuidadosamente retirado do molde com as próprias mãos, puxando-o até que se desprenda. Sua superfície convexa que estava em contato com o gesso deve agora ser limpa e receber algumas pinceladas de resina. Deve-se esperar secar por um dia. Em seguida deve-se preparar gesso líquido suficiente para preencher a parte côncava do contra-molde de fibra de vidro e resina, para que o contra-molde adquira rigidez. Pois a parte do contra-molde que será utilizada neste experimento é a convexa. É sobre esta superfície convexa que será esticado o papel laminado.

O contra-molde pode ser visto na Fig. 4, onde mostramos a superfície convexa que deve ser muito bem polida para dar melhores resultados neste experimento.

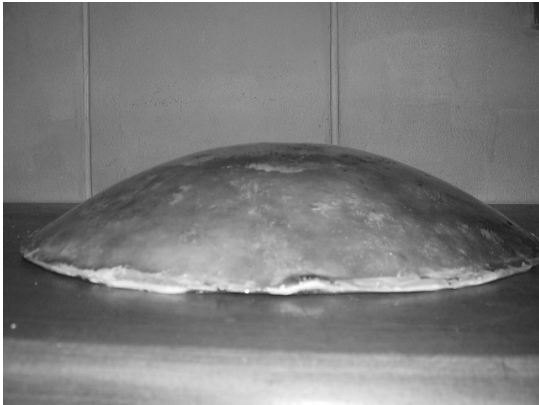

**Fig. 4:** Contra-molde parabólico feito de fibra de vidro e resina.

## 5. Espelhamento

Na construção de um espelho é necessário determinar o processo de espelhamento. Ou seja, o processo de aderir à superfície lisa do espelho uma camada refletora de alumínio ou prata. Os processos mais comuns são a deposição de um filme de prata ou de alumínio por evaporação térmica em uma câmara de vácuo, ou a deposição do filme por solução de prata. O processo que utilizamos é diferente dos usuais, já que usamos somente papel laminado para fazer o espelhamento.

Uma folha de papel laminado retangular de aproximadamente 50 cm x 40 cm é constituída por três camadas. Uma camada fina de plástico transparente, uma camada fina de filme de prata aderida a este plástico, dando origem a um plástico filmado e, por fim, uma camada de papel mais grossa aderida ao plástico filmado. Um outro tipo de papel laminado aparentemente idêntico ao descrito, mas diferindo deste quanto ao processo de fabricação, é aquele constituído por um papel filmado aderido a um plástico fino e transparente. Este tipo não é adequado para o nosso processo de aluminização, uma vez que o filme de prata não está depositado no plástico, mas sim no papel.

Para poder fazer cortes em formas de raias no papel laminado foram feitas quatro séries de traços na parte de papel do papel laminado do primeiro tipo, conforme esquema mostrado na Fig. 5. Os traços primários, ou os da primeira série, são dois traços contínuos, cada um dividindo o papel laminado ao meio e cruzando-se perpendicularmente na região central do papel. Os traços secundários, ou os da segunda série, são dois círculos contínuos e concêntricos no ponto de cruzamento dos traços primários com diâmetros diferentes. O diâmetro de um dos círculos é tal que este círculo seja tangente a uma das bordas do papel laminado e totalmente contido nele. O diâmetro do outro círculo pode ser entre um a três centímetros menor do que o primeiro diâmetro. Os traços terciários, ou os da terceira série, são vários traços contínuos passando pelo ponto de cruzamento dos traços primários e com extremidades em pontos diametralmente opostos do círculo maior. O conjunto destes traços divide este círculo em fatias iguais com um arco interno, devido ao círculo menor, e um arco externo com dois centímetros e meio aproximadamente, devido ao círculo maior. E, por fim, os traços quaternários, ou os da quarta série, são aqueles traços contínuos paralelos aos traços terciários e com extremidades nos traços secundários, ou círculos, onde uma dessas extremidades pertence à parte central do arco interno da fatia correspondente.

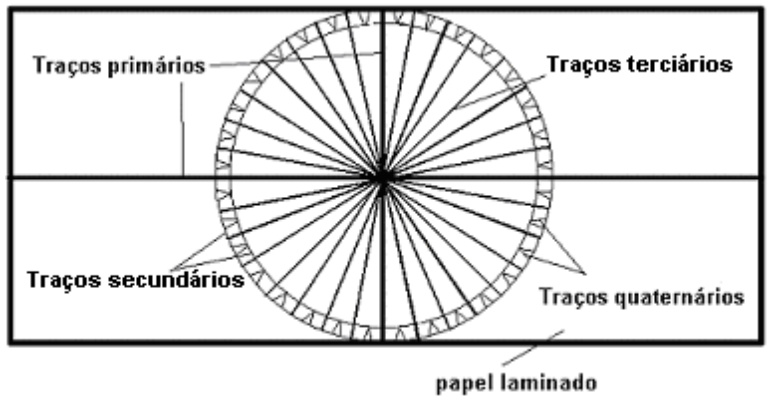

**Fig. 5:** Traços no papel laminado para a realização do corte em forma de raias.

Em seguida recortamos este papel traçado de modo que ele ficou com um disco central e muitas fitas nas extremidades. Para isso, o primeiro procedimento foi recortar o papel traçado com uma tesoura acompanhando o círculo maior, resultando assim em um disco de papel laminado e traçado. O segundo procedimento foi recortar com uma tesoura este disco, acompanhando exatamente os traços quaternários, resultando em raias de dois centímetros e meio aproximadamente, com um disco central delimitado pelo círculo menor.

Fitas adesivas foram colocadas ao redor do disco acompanhando longitudinalmente o círculo menor de modo que metade da fita grudasse em três a cinco raias e outra metade grudasse no disco central. Estas fitas são importantes para que o disco de plástico filmado não seja rasgado durante o processo de esticamento que será descrito adiante. Depois que contornamos o disco grudando fita adesiva, colocamos este papel recortado em água corrente. Esfregando com os dedos a parte de papel do disco vamos retirando o papel gradativamente, restando apenas o plástico filmado com películas de papel aderidas. Estendendo este plástico em um varal, espera-se que ele seque à sombra. A Fig. 6 mostra uma fotografia de como ficou o papel laminado depois de aplicar os procedimentos descritos.

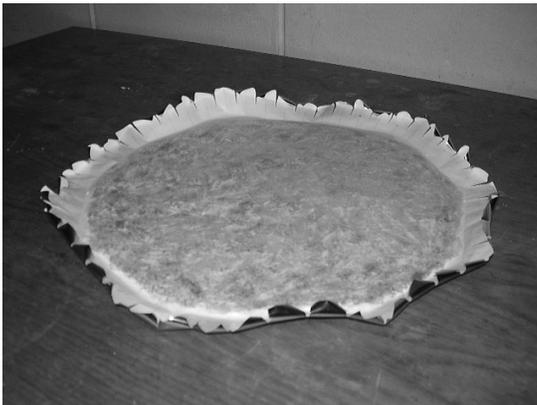

**Fig. 6:** Papel laminado depois de recortado e retirada a parte de papel.

O recipiente deve servir de apoio auxiliar ao colocá-lo sobre um local firme com sua base voltada para cima. A base do contra-molde deve ser colocada sobre a base do recipiente. Antes de colocarmos o plástico com filme de prata sobre o contra-molde, esparramamos um pouco de óleo de cozinha sobre o contra-molde para que o plástico não fique em contato direto com ele. O óleo ajuda a diminuir as forças de coesão entre a superfície de plástico e a superfície do contra-molde, já que estas superfícies estão muito bem ajustadas uma à outra.

Então, depois de seco, o papel assim preparado é colocado sobre o contra-molde de fibra de vidro e resina, com as películas de papel voltadas para cima para o papel ser esticado com fita adesiva. Uma extremidade da fita adesiva é grudada em uma raia atingindo a fita adesiva colocada anteriormente no disco de plástico até ultrapassá-lo em aproximadamente meio centímetro, recaindo sobre o disco de plástico. A outra extremidade é grudada no recipiente atingindo a parte interna do mesmo para poder esticar o plástico com filme de prata. O processo de esticamento é simples. Depois que todas as raias estiverem presas no recipiente com as fitas adesivas, cada fita adesiva é desgrudada do recipiente, esticada na direção oposta ao disco central do plástico e grudada novamente no recipiente. O óleo que atinge a parte de baixo das raias deve ser removido quando se desgruda a fita para esticá-la. Se este óleo invadir a parte aderente da fita adesiva, esta perde sua atividade. Este procedimento de esticamento é realizado fita por fita até que toda a área do plástico com filme de prata esteja sem nenhuma ruga, ou seja, totalmente esticado e deformado tomando a forma do contra-molde parabólico.

## 6. Espelho Parabólico

Para a fabricação do espelho parabólico prepara-se um gesso pastoso para colocar sobre este plástico esticado com fita adesiva no contra-molde. Uma quantidade de gesso deve ser preparada de modo que cubra quase toda a área do disco de plástico esticado e de modo que também tenha uma certa espessura quando secar.

Para preparar esta consistência de gesso basta ir adicionando pouca água ao pó de gesso e ao mesmo tempo ir misturando. E então, depois de misturar até ganhar homogeneidade, despeja-se sobre o plástico. Espera-se secar. Quando estiver seco, as fitas adesivas que foram usadas no processo de esticamento devem ser rompidas. Deve-se realizar um torque com as mãos no gesso já seco, forçando-o a se soltar do contra-molde. Depois que ele se soltar será observado que o plástico grudou perfeitamente no gesso, constituindo uma superfície espelhada e parabólica. A Fig. 7 mostra o espelho parabólico rugoso uma vez que a superfície do contramolde não foi polida.

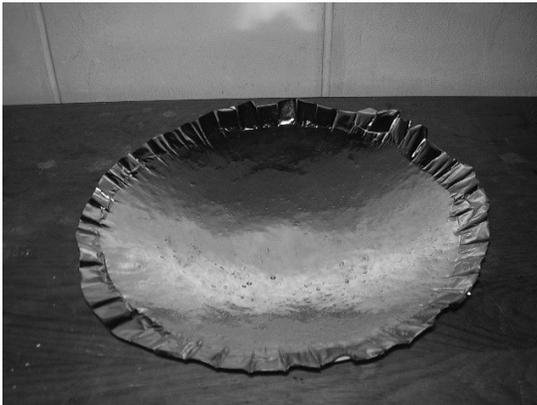

**Fig. 7:** Espelho parabólico rugoso, pois a superfície do contra-molde não foi polida.

A adesão do plástico filmado ao gesso se deve ao fato de que, durante o processo da retirada de papel do papel laminado, uma pequena camada de películas de papel ainda permaneceu sobre o filme de prata. E é esta camada que aderiu ao gesso. Pois foi realizada uma experiência onde esta camada foi totalmente retirada. O resultado foi que o gesso solidificado não grudou no plástico, produzindo uma superfície parabólica muito lisa que poderia receber um filme metálico por outros métodos, tais como: ou evaporação térmica ou solução de prata. Para aumentar o espelhamento poderia ser pulverizada tinta prata sobre a superfície do plástico, sem a camada de fibra de papel. Para aumentar a refletividade a superfície convexa do contra-molde poderia ser muito bem polida.

A Fig. 8 mostra dois espelhos que foram produzidos sobre uma superfície côncava muito bem polida, embora não parabólica e sim esférica.

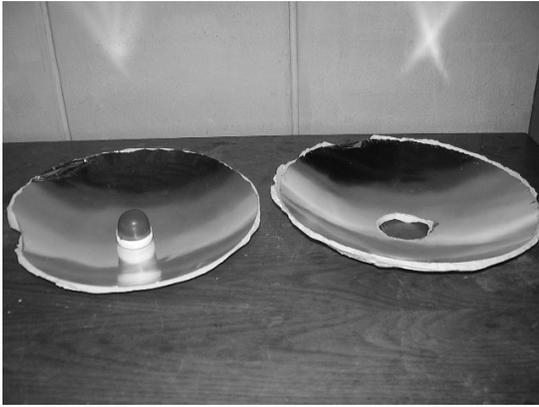

**Fig. 8:** Espelho côncavo produzido com papel laminado esticado sobre uma superfície polida.

A refletividade do espelho neste caso é muito maior que no caso anterior. Como o gesso fossiliza a superfície, a camada refletora do espelho terá a forma da superfície na qual o plástico filmado foi esticado.

## 7. Conclusão

O procedimento descrito aqui para a construção de espelhos parabólicos não foi encontrado na literatura. O nosso pioneirismo se encontra no fato de termos solidificado o formato parabólico dos líquidos em rotação e ao construirmos um espelho parabólico com procedimento e materiais não usuais, de baixo custo e facilmente encontráveis no comércio.

Obviamente são possíveis melhoramentos nos procedimentos aqui descritos. Também podem ser explorados outros resultados nas etapas que descrevemos. Como possíveis exemplos citamos a quantificação das variáveis no sistema girante, o polimento do contra-molde de fibra de vidro e resina, o aumento da refletividade do espelho com tinta prata, a deposição de uma camada refletora na superfície lisa do gesso por evaporação térmica ou por solução de prata, etc. E também a utilização deste método não se restringe a espelhos esféricos, mas qualquer aplicação de calotas para experimentos diversos, tais como oscilação pendular de esferas amortizadas por atrito, movimentos concêntricos em espiral em calotas, centrifugação de líquidos, etc. Não pretendemos esgotar o assunto com este trabalho, mas sim mostrar várias possibilidades inovadores que ele apenas inicia. Nosso objetivo aqui foi apenas o de apresentar um procedimento novo para a construção de espelhos parabólicos utilizando materiais de baixo custo. Isto permite que este procedimento seja explorado por professores e alunos de ensino médio e universitário.

## 8. Agradecimentos

## 9. Referências